\documentclass[lettersize,journal]{IEEEtran}
\usepackage[table]{xcolor}   

\usepackage{amsmath}
\usepackage{amssymb}
\usepackage{amsthm}
\usepackage{tikz}
\usepackage{amsfonts,xspace,comment}
\usepackage{algorithmic}
\usepackage{textcomp}
\usepackage{stfloats}
\usepackage{url}
\usepackage{algorithm2e}
\usepackage{verbatim}
\usepackage{tabularx}
\usepackage{makecell}
\def\BibTeX{{\rm B\kern-.05em{\sc i\kern-.025em b}\kern-.08em
    T\kern-.1667em\lower.7ex\hbox{E}\kern-.125emX}}
\usepackage{balance}
\usepackage{array}
\usepackage{graphicx}

\usepackage{soul}
\usepackage[hidelinks]{hyperref}
\usepackage{multirow}
\usepackage{colortbl}
\usepackage{hhline}
\usepackage{booktabs}
\usepackage{enumitem}
\usepackage{float}
\usepackage{caption}
\usepackage{fancyhdr}
\usepackage{orcidlink}
\usepackage{ragged2e}
\usepackage{subcaption}

\usepackage{orcidlink}

\newcommand{\ours}{{AdvChar}\xspace}
\def\BibTeX{{\rm B\kern-.05em{\sc i\kern-.025em b}\kern-.08em
    T\kern-.1667em\lower.7ex\hbox{E}\kern-.125emX}}

\newcommand*\cibbox[1]{\tikz[baseline=(char.base)]{\node[shape=rectangle,fill=gray!30!white,text=black,inner sep=1.5pt] (char) {#1};}}
\newcommand{\BfPara}[1]{{\vspace{0.5ex}\noindent\bf#1.}\xspace}
\newcommand{\UlBfPara}[1]{{\vspace{1ex}\noindent\bf\cibbox{#1.}}\xspace}

\newcommand{\eg}{{\em e.g.,}\xspace}
\newcommand{\ie}{{\em i.e.,}\xspace}
\newcommand{\etc}{{\em etc.}\xspace}

\newcommand{\ed}[1]{\textcolor{black}{#1}}

\setul{0.8pt}{}
\hypersetup{
	colorlinks=true,
	urlcolor=black!70!black,
	linkcolor=red!70!black,
	citecolor=purple,
} 

\begin{document}

\title{Attacking interpretable NLP systems}

\author{Eldor~Abdukhamidov \orcidlink{0000-0001-8530-9477},
Tamer~Abuhmed~\orcidlink{0000-0001-9232-4843},
Joanna~C.~S.~Santos~\orcidlink{0000-0001-8743-2516}, and
        Mohammed~Abuhamad~\orcidlink{0000-0002-3368-6024}
        
        \IEEEcompsocitemizethanks{\IEEEcompsocthanksitem Eldor Abdukhamidov and Tamer Abuhmed are with the Department of Computer Science and Engineering, Sungkyunkwan University, Suwon, South Korea.\protect
        (E-mail: abdukhamidov@skku.edu, tamer@skku.edu). 
        Joanna C. S. Santos is with the Department of Computer Science and Engineering, University of Notre Dame, Notre Dame, USA.
        (joannacss@nd.edu)
        Mohammed Abuhamad is with the Department of Computer Science, Loyola University, Chicago, USA.\protect
        (E-mail: mabuhamad@luc.edu).\\     }
}

\markboth{ }%
{Shell \MakeLowercase{\textit{et al.}}: A Sample Article Using IEEEtran.cls for IEEE Journals}

\maketitle

\begin{abstract}
Studies have shown that machine learning systems are vulnerable to adversarial examples in theory and practice. Where previous attacks have focused mainly on visual models that exploit the difference between human and machine perception, text-based models have also fallen victim to these attacks. However, these attacks often fail to maintain the semantic meaning of the text and similarity. \ed{This paper introduces \ours{}, a black-box attack on Interpretable Natural Language Processing Systems, designed to mislead the classifier while keeping the interpretation similar to benign inputs, thus exploiting trust in system transparency.}
\ours{} achieves this by making less noticeable modifications to text input, forcing the deep learning classifier to make incorrect predictions and preserve the original interpretation. 
We use an interpretation-focused scoring approach to determine the most critical tokens that, when changed, can cause the classifier to misclassify the input. 
We apply simple character-level modifications to measure the importance of tokens, minimizing the difference between the original and new text while generating adversarial interpretations similar to benign ones. 
We thoroughly evaluated \ours{} by testing it against seven NLP models and three interpretation models using benchmark datasets for the classification task. Our experiments show that \ours{} can significantly reduce the prediction accuracy of current deep learning models by altering just two characters on average in input samples.
\end{abstract}


\section{Introduction}

Deep learning models, particularly in Natural Language Processing (NLP), have revolutionized how machines understand and interact with human language. These advancements have enabled various applications, from simple spellcheck and keyword search to complex tasks such as sentiment analysis~\cite{wankhade2022survey}, machine translation~\cite{dabre2020survey}, and chatbot interactions~\cite{ni2023recent}. The integration of NLP into our daily digital interactions, such as through search engines, virtual assistants, and recommendation systems, highlights its importance. However, these models are shown to be susceptible to adversarial attacks \cite{NEURIPS2023_d2b752ed}.

Adversarial attacks in NLP, which involve careful manipulations of input data leading to incorrect model outputs, are a growing concern. These attacks are especially stealthy because of the complex nature of human language, which is filled with idioms, metaphors, and context-dependent meanings~\cite{zhang2020adversarial}. It is essential to rigorously research and develop effective methods to counter these adversarial attacks, given the growing integration of such models in various NLP applications across different sectors.
For example, adversarial attacks on AI systems for social media moderation, opinion analysis, customer reviews, or market trends could result in failures to identify harmful content, misinterpretation of benign posts, and misleading analyses, ultimately leading to poor decisions~\cite{alsmadi2023adversarial}. 

\begin{figure}[t]
    \centering
    \captionsetup{justification=justified}
    \includegraphics[width=\linewidth]{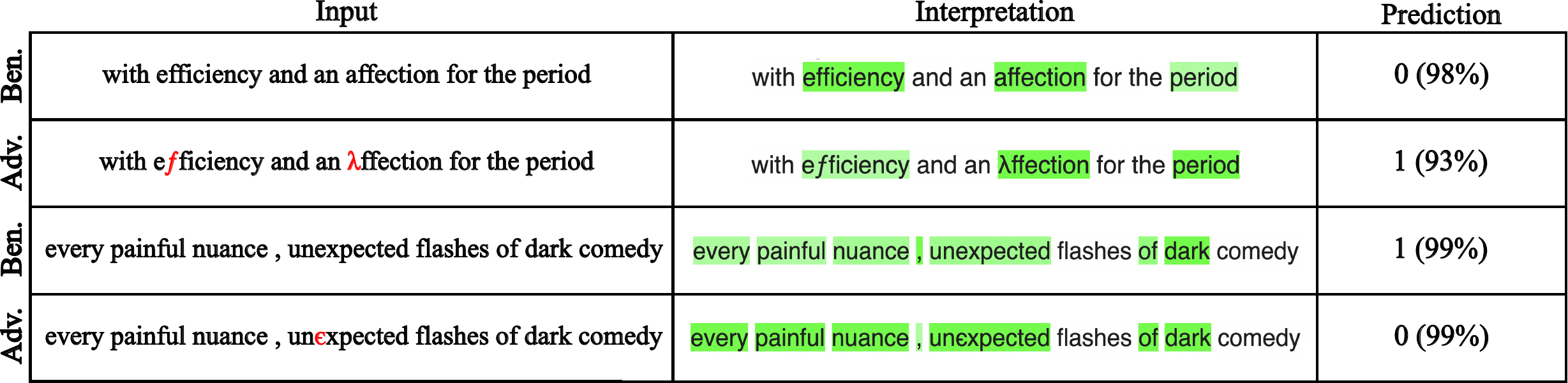}
    \caption{\ed{Example texts comparing a benign sample and a sample subject to our proposed attack, along with their corresponding interpretations based on LIME interpreter. Both inputs generate similar interpretations regardless of differences.}}
    \label{fig:intro_images}
\end{figure}

By coupling NLP classifiers with interpreters, \ie creating
Interpretable NLP Systems (INLPS), adversarial manipulations can be recognized by an observer (see \autoref{fig:intro_images}).
INLPS offer transparency in the decision-making process, which is especially crucial for applications where understanding the reasoning behind decisions is as important as the decisions themselves. 
Although this transparency is designed to promote trust and accountability, it can inadvertently introduce vulnerabilities, offering potential attackers a roadmap to manipulate the outcomes. 
This vulnerability is not just a theoretical concern but a practical challenge that needs to be addressed to ensure the reliability and safety of AI applications. 
In highlighting the vulnerabilities introduced by INLPS, our research goes a step further by demonstrating these theoretical concerns in a practical context. 
\ed{Our attack, \ours{}, targets Interpretable Natural Language Processing Systems (INLPS), which integrate NLP classifiers with interpreters to deliver predictions and explanations. The attack has two objectives: (1) to mislead the classifier into misclassifying the input, and (2) to ensure that the interpreter produces an explanation similar to that of a benign input.} 
We focus on text classification tasks using large language models (LLMs) and diverse interpretation methods.
By changing certain words at the character level, \ours{} is designed to mislead both the primary NLP classifier and its interpreter. 
\ours{} demonstrates that the strengths of INLPS can also become weaknesses (as shown in  \autoref{fig:intro_images}), emphasizing the need for robust countermeasures.

\ed{We evaluate our attack using three distinct datasets: the Stanford Sentiment Treebank (SST-2), AG News, and Yahoo Answers datasets. SST-2 (with binary sentiment classification) serves as a basic test for our attack, while AG News (with four-category news topic classification) and Yahoo Answers (with ten-category question classification) provide more complex scenarios to assess the effectiveness of the attack.}

We selected a diverse range of LLM models, \ie GPT-2, BERT, DistilBERT, Electra, CANINE, FNet, and XLM-R, to provide a comprehensive analysis across these models.
Furthermore, our study incorporates three interpretation models, \ie SHAP \cite{NIPS2017_7062}, Saliency Maps \cite{selvaraju2017grad}, and LIME \cite{ribeiro2016should}, each providing unique insights into the model decision-making process. 
\ed{Our findings show that the attack reaches a success rate of 79\% in the AG News dataset using LIME interpreter with  CANINE model. In the SST-2 and Yahoo Answers datasets, it achieves a success rate of 79\% (with Saliency Map interpreter and BERT model) and 80\% (with LIME interpreter and BERT model), respectively. The effectiveness of the attack varies by models and interpreters, with LIME generally outperforming others, especially on complex tasks.}

\BfPara{Contributions} We summarize our contributions as follows:
\begin{itemize}[leftmargin=1em]
    \item We propose a stealthy and query-efficient black-box attack targeting INLPS. We evaluate the attack performance against seven classifiers when coupled with three different interpreters using {three} datasets. The evaluations show that the proposed attack achieves considerable success rates, highlighting the attack's effectiveness in various scenarios.
    \item We investigate the transferability of adversarial examples across various classifiers and interpreters to assess whether adversarial samples designed for a certain model and interpreter can expose vulnerabilities in other models and interpreters, highlighting areas of concern in the field.
    \item We perform a comparative analysis to examine several factors (\ie input length, amount of perturbation, \etc) that might influence the performance of the attack.
\end{itemize}

\BfPara{\textbf{Organization}} Our paper is organized as follows:
\autoref{sec:related} reviews related research studies;
 \autoref{sec:notations} describes the notations and terms used in the paper; 
 \autoref{sec:notations} presents the proposed attack and its underlying mechanisms;  
 \autoref{sec:evaluation} provides the results of the attack effectiveness, robustness, and transferability against LLMs and interpretation models; 
 \autoref{sec:discussion} explains the existing limitations and future work; and \autoref{sec:conc} concludes the paper.

\section{Related Work} \label{sec:related}
The evolution of NLP techniques has advanced significantly with the emergence of deep learning-based LLMs.
Despite their strength, deep learning models have demonstrated certain vulnerabilities, particularly when faced with adversarial perturbations.
Several studies \cite{bao2021defending, ren2019generating,10381813,10562350} have highlighted how these models, despite their complexity, can be deceived by carefully crafted perturbations in the input.

NLP presents a unique challenge compared to other domains, such as computer vision, due to the nature of its data. Textual data is inherently discrete, contrasting with images' continuous nature. For example, a sentence like `\textit{I do not dislike this movie}' in NLP can be challenging to interpret. While `\textit{dislike}' generally suggests something negative, the addition of `\textit{do not}' shifts the meaning to positive or neutral. 
This highlights how text meaning can change significantly based on word arrangement, emphasizing its discrete nature. \ed{In contrast, CNN-based image classifiers typically show robustness against minor modifications like changing one or a few pixels, unless these modifications are carefully designed as adversarial perturbations~\cite{abdukhamidov2021advedge,abdukhamidov2023hardening, abdukhamidov2022black, abdukhamidov2024singleadv}. This fundamental distinction highlights why adversarial attacks on textual data are inherently challenging. Even minor textual perturbations can substantially change the intended meaning and model predictions, making subtle adversarial modifications notably harder to achieve \cite{bhambri2019survey,chen2020seqvat,li2021tianyu, cheng2019robust}.}

Using word embeddings/representations in NLP adds complexity to the problem of adversarial attacks. Word embeddings convert words into continuous vector spaces, which can be modified to introduce perturbations. By slightly adjusting the vectors associated with words, attackers can mislead a model into interpreting the input differently, resulting in incorrect outputs. This aspect of adversarial attacks, especially in the context of word embeddings, has been extensively explored in research studies such as \cite{goyal2023survey, li2023efficiently, liu2020joint}.

Various tasks in NLP range from basic predictive tasks, \eg sentiment analysis \cite{10.1145/3649451} and text classification \cite{fields2024survey}, to more complex generative tasks, \eg  machine translation \cite{NEURIPS2023_11c7f1dd}, and question answering \cite{Li_Fan_Gu_Li_Duan_Dong_Liu_Wang_2024}. This diversity of tasks also means that adversarial attacks can occur in different scenarios and threat models. Each task has its unique vulnerabilities that attackers exploit with creativity. In our work, we focus mainly on text classification tasks.

\section{Notations and Definitions} \label{sec:notations}

This section introduces the notation and definitions used throughout the paper to describe our attack and its associated concepts. The symbols and notation used in this
paper are summarized in \autoref{tab:notations}.
The samples are denoted as pairs ($x$,$c$). Specifically, $[x_1,x_2,x_3,...,x_n] = x \subset X$, where $x$ refers to an input text sequence of $n$ tokens. 
These tokens can be words or characters, depending on the specific model used. Meanwhile, {\( c = 1, ..., M \)} represents a label of $M$ classes. The deep learning model is illustrated as \( F: X \rightarrow C \), which maps the input set to the associated labels. 

\begin{table}[!ht]
\centering
\caption{Notations used in the paper.}
\rowcolors{2}{gray!15}{white}  
\begin{tabular}{ c l }
\toprule
\textbf{Symbol} & \textbf{Description} \\
\midrule
\( x \) & Original text sequence. \\
\( x' \) & Adversarial text sequence generated by the attack. \\
\( x_i \) & \( i^{th} \) token in the sequence \( x \). \\
\( c \) & Output label of benign input text. \\
\( c' \) & Output label of adversarial input text. \\
\( g \) & Benign interpretation map. \\
\( g' \) & Adversarial interpretation map. \\
\( F \) & Target deep learning classifier. \\
\( G \) & Target interpreter. \\
\( T \) & General notation for token transformation function. \\
\( S \) & Similarity function. \\
\( m \) & Number of top important tokens to be perturbed. \\
\( T_{\text{sub}}(x_i) \) & Substitution transformation applied to token \( x_i \). \\
\bottomrule
\end{tabular}
\label{tab:notations}
\end{table}

\BfPara{Classifier} {In this study}, our primary focus is on predictive tasks, such as text classification, in which a deep learning model $F$ classifies an input $x$ into one of the predefined classes \( C \), with the result being $F(x)=c$ that belongs to $C$. 

\BfPara{Interpreter} There exist two distinct approaches to improving the interpretability of deep learning models: one involves building models with inherent comprehensibility, while the other involves the adoption of post hoc explanation techniques. Using post hoc techniques does not require any modifications to the model's architecture or configuration, and therefore, they are often more desirable.
Our experiments primarily focus on post hoc explanations, which explain how a model works after it has made a decision, focusing specifically on individual examples. We examine how the model, which we denote as \( F \), processes a given input \( x \) and how it makes a decision based on that input, referred to as \( F(x) \). We interpret the connections between \( x \) and \( F(x) \) as interpretation maps. We use an interpreter, \( G \), to create a map \(g = G(x, F)\) for each input \( x \) using the model \( F \). 
In this map, \( g \), each element \( g[i] \) highlights the importance of the \( i \)-th component of the input \( x \) in generating the output \( F(x) \).

\BfPara{Threat Model} 
{This work assumes a black-box scenario: the adversary does not have direct knowledge of the inner workings of the classifier \( F \) and/or the interpreter \( G \), including their structures and parameters. 
This assumption represents scenarios in which the internal details of the system are hidden from potential adversaries, which is practical since modern deep learning models are often used as a service that takes input from a user and produces a related output.}
An adversary/user interacts with the system by providing inputs to the classifier \( F \) and receiving the corresponding outputs. In a typical use case, the user submits an input \( x \), the classifier processes this input, and the user receives the output \( c = F(x) \). Additionally, the interpreter \( G \) can be queried (by the system/observer) to provide explanations \( g = G(x, F) \) for the classifier's output.

\BfPara{\ul{Adversarial Capabilities and Constraints}}
The adversary has the following capabilities and constraints:
    \begin{itemize}[leftmargin=*]
        \item {Query the classifier \( F \) and interpreter \( G \) with any input \( x \).}
        \item {Observe the classifier's output \( c = F(x) \).}
        \item {Observe the interpreter's output \( g = G(x, F) \).}
\item {Does not know/access training data or methods for \( F \) or \( G \).
\item Can not access the parameters or architecture of \( F \) or \( G \).
        \item Can not modify the internal components of \( F \) or \( G \).}
    \end{itemize}

{\BfPara{\ul{Adversarial Objectives}}}
The adversary aims to design the inputs \( x' \) such that the classifier's output \( c' = F(x') \) is incorrect, while \( x' \) appears similar to a benign input.
The adversary seeks to maintain the original explanations provided by the interpreter \( G \) to hide the true nature of the classifier's decision-making process or to mislead users regarding the classifier's reliability.

Taking into account these scenarios, capabilities, constraints, and objectives, this work evaluates the robustness of the system against adversarial attacks.

\section{Methodology} \label{sec:methods}
This section provides an overview of the attack and its execution on three different types of interpreters.

\subsection{Attack Formulation}
\ed{In \ours{}, the interpretable model is considered as a central target of the attack. It is used to guide character-level perturbations and preserve overall interpretation, ensuring that adversarial input remains undetectable even under explanation-based inspection.} Adversarial attacks generate an adversarial sample by producing a perturbation \( \Delta x \) within a specific threshold $\epsilon$. 
\ours considers more constraints since the target systems employ interpretable models for the decision-making process. The goal of \ours{} is twofold: to trick the classifier and to ensure that the interpreter produces interpretation maps that resemble those of benign cases:

\begin{enumerate}[leftmargin=1.5em]
    \item Ensuring model misclassification, \ie $F(x')$ $\rightarrow$ $c'$, $c'$ $\neq$ $c$.
    \begin{equation} \label{eq:advFormula}
    F(x) \rightarrow c, \quad F(x') \rightarrow c',  \quad c' \neq c
\end{equation}
    \item Triggering an interpreter $G$ to generate target interpretation maps, \ie $G(x'; F) \rightarrow g': g' \approx g$, where $G(x; F) = g$.
\begin{equation} \label{eq:advFormula}
G(x, F) \rightarrow g, \quad G(x', F)  \rightarrow g',  \quad g' \approx g
\end{equation}
    \item The variation between $x$ and $x'$, denoted as $\Delta(x, x')$, should not be noticeable.
\end{enumerate}

\ed{Unlike traditional attacks, such as TextFooler \cite{jin2020bert} or PWWS \cite{ren2019generating} that score word importance based on prediction confidence shifts, \ours{} leverages the output of interpretation models to identify influential characters. This enables the attack to simultaneously influence both the classifier's decision and its explanation, which is crucial for targeting INLPS.}

To make the perturbation small enough to be unnoticeable to humans, we implement \ours in the character level to achieve an adversarial attack. In other words, \ours generates adversarial characters visually similar to the target character but unfamiliar to the target NLP model. The attack is universal, meaning it can be applied to any NLP model, and it generates adversarial samples that produce interpretations similar to the benign input, making it challenging for humans to detect the adversarial samples.

Our adversarial strategy involves the following steps:

\UlBfPara{1) Token Importance Evaluation} 
\ed{Using an interpreter output from the INLPS, such as visualized attention or color intensity maps, we estimate each token's importance without relying on model probabilities or gradients. These outputs are converted into numerical importance scores, providing a practical and realistic way to identify influential tokens. This enables \ours{} to function effectively even in black-box API scenarios.}
Therefore, we can derive an importance score for each element (token) $\mathcal{I}(t_i)$. These scores can be organized in ascending order to target the most influential elements.

Let us denote the set of elements (\ie tokens) from the input $x_i$ as 
\(x_i = \{t_1, t_2, ..., t_n\} \).
Upon computing their respective importance scores, we can re-arrange them into a sorted set \( x_i^* \) such that:
\( x_i^* = \{t_1^*, t_2^*, ..., t_n^*\} \) where
\( \mathcal{I}(t_1^*) \geq \mathcal{I}(t_2^*) \geq ... \geq \mathcal{I}(t_n^*) \).
In this representation, \( t_1^* \) is the element with the highest importance score and \( t_n' \) has the least. This ordered set facilitates a structured approach, allowing us to prioritize and perturb the most influential elements first.

Since interpreters have varied output ranges, importance scores are adjusted to typically lie between 0 and 1. This ensures that no particular element or score disproportionately influences the model due to its magnitude. Normalization is performed as follows:
$\mathcal{I}(t_i) = [{\mathcal{I}(t_i) - \min(\mathcal{I}(x_i))}]\div[{\max(\mathcal{I}(x_i)) - \min(\mathcal{I}(x_i))}]$.

\UlBfPara{2) Adversarial Perturbation} Starting with the essential token, we introduce slight modifications. These perturbations are crafted to mislead the classifier while the token remains semantically readable and keeps its importance.
Given:
\begin{itemize}
    \item[(1)] \( C(t_i) \) as the set of characters in token \( t_i \)
    \item[(2)] \( R(t_i, \zeta) \) as a function that replaces character \( \zeta \) in input token \( t_i \) with a new character
    \item [(3)] \( L(F, x_i, c) \) as the classification loss for input \( x_i \) and true label \( c \) using classifier \( F \)
    \item[(4)] \( S(x_i, x'_i) \) as a similarity function that measures the difference in the order of importance of the tokens before and after perturbation for the benign sample \( x_i \).
\end{itemize}
The token perturbed by our attack can be represented as:
$$t_i' = R (t_i, \zeta_r) ~\text{s.t.} ~ \zeta_r \in C(t_i)$$

\ed{We define the similarity function as:}
\begin{align*}
    S(x_i, x'_i) = \|\text{argsort}_{t_i \in x_i} [\mathcal{I}(t_i, F, G)] - \\\text{argsort}_{t_i' \in x'_i} [\mathcal{I}(t_i', F, G)]\| 
    \quad\text{s.t.}\quad S(x_i, x'_i) \leq \theta,
\end{align*}
\ed{where $\text{argsort}_{t_i \in x_i} [\mathcal{I}(t_i, F, G)]$ represents the indices of tokens in $x_i$ sorted by their importance scores $\mathcal{I}(t_i, F, G)$ in descending order. Here, the threshold $\theta$ limits permissible changes in token ordering after perturbation. \ours continues to perturb tokens following this order until the perturbation threshold $\theta$ is reached or the attack succeeds, \ie $F(x') \neq c$.}

\UlBfPara{3) Iterative Attack} If the classifier is not deceived by perturbing the essential token, \ours{} keeps the modification and proceeds to the next token in the order of importance. \ed{This iterative process continues until the classifier is fooled, the similarity exceeds the perturbation threshold $\epsilon$, or all tokens have been considered.}
In this process, all tokens in the input text are considered for perturbation to ensure comprehensive coverage and maximize the chances of deceiving the classifier.

The details of the algorithm are presented in \autoref{alg:attack_algorithm}.

\RestyleAlgo{ruled}
\SetKwComment{Comment}{/* }{ */}
\newcommand\mysim{\mathrel{\stackrel{\makebox[0pt]{\mbox{\normalfont\tiny rand}}}{\sim}}}

\begin{algorithm}[t]
\scriptsize
\caption{\ours{} attack's main algorithm}\label{alg:algorithm_pertb}
\label{alg:attack_algorithm}
\KwData{Classifier $F$, interpreter $G$, input text $x$, similarity function $S$, benign category $c$, interpretation map $m$, perturbation threshold $\epsilon$.} 

\KwResult{Adversarial text $x'$}
\textbf{Initialization:} Setting $x' \leftarrow x$\;
\Begin{

    $g \leftarrow{} G(x, F)$ \Comment*[r]{extract interpretation map}   
    $scores \leftarrow Convert(g)$ \Comment*[r]{convert map into importance scores}

    $x_{ordered} \leftarrow Sort(x_1, x_2, x_3, ..., x_n, scores_{x_i})$ \Comment*[r]{sort tokens based on importance score}
    
    \For{$x_i$ in $x_{ordered}$}{
        $x_i' \leftarrow SubstituteChar(x_i)$ \Comment*[r]{replace a character in a word}
        $x' \leftarrow \; replace \; x_i  \; with \; x_i' \; in \; x'$ \;
        \uIf{$S(x, x') > \epsilon$}{
            \Return{None} \Comment*[r]{Perturbation threshold is reached}
        }
        \uElseIf{$F(x') \neq c$}{
            \Return{$x'$} \Comment*[r]{Solution is found}
        }
        \Else {
            \textbf{continue}
        }
    }

    \Return{None} \Comment*[r]{Attack failed}
}
\end{algorithm}

\subsection{NLP Classification Models and Interpreters}

\BfPara{Classification Models} Diverse NLP models are employed to test our attack. Our experiment includes GPT-2 \cite{radford2019language}, BERT \cite{kenton2019bert}, DistilBERT \cite{sanh2019distilbert}, Electra \cite{clark2020electra}, CANINE \cite{clark2022canine}, FNet \cite{lee2021fnet}, and XLM-R \cite{conneau2019unsupervised} models. The selected models offer a comprehensive overview of the advancements in NLP models over recent years. They collectively represent a range of architectures and methodologies, from lightweight and efficient versions to those optimized for multilingual tasks or generative capabilities. By incorporating such a diverse set of models, the experiment aims to capture a holistic understanding of the capabilities and performance variations in the modern NLP domain. While some models are designed for classification tasks, others have been fine-tuned and transformed effectively as classification models. The performance of the models on {three} datasets are provided in \autoref{tab:nlp_model_performance}. 

\begin{table}[]
\centering
{\color{black}
\caption{\ed{Performance of NLP models on three different datasets. The performance is provided in terms of accuracy.}}
\label{tab:nlp_model_performance}
\centering
\resizebox{0.9\linewidth}{!}{%
\begin{tabular}{ccccccc}
\toprule
\multicolumn{1}{c|}{\textbf{GPT-2}} & \multicolumn{1}{c|}{\textbf{BERT}} & \multicolumn{1}{c|}{\textbf{DistilBERT}} & \multicolumn{1}{c|}{\textbf{Electra}} & \multicolumn{1}{c|}{\textbf{CANINE}} & \multicolumn{1}{c}{\textbf{FNet}} & \textbf{XLM-R} \\ \midrule
\multicolumn{7}{c}{SST-2}                                                                                                                                                                \\ 
\multicolumn{1}{c|}{0.92}  & \multicolumn{1}{c|}{0.91} & \multicolumn{1}{c|}{0.97}       & \multicolumn{1}{c|}{0.95}    & \multicolumn{1}{c|}{0.85}   & \multicolumn{1}{c}{0.89} & 0.93  \\ \midrule
\multicolumn{7}{c}{AG}                                                                                                                                                              \\ 
\multicolumn{1}{c|}{0.94}  & \multicolumn{1}{c|}{0.95} & \multicolumn{1}{c|}{0.95}       & \multicolumn{1}{c|}{0.94}    & \multicolumn{1}{c|}{0.91}   & \multicolumn{1}{c}{0.91} & 0.92  \\ \midrule

\multicolumn{7}{c}{Yahoo Answers}                                                                                                                                                              \\ 
\multicolumn{1}{c|}{0.90}  & \multicolumn{1}{c|}{0.91} & \multicolumn{1}{c|}{0.89}       & \multicolumn{1}{c|}{0.89}    & \multicolumn{1}{c|}{0.87}   & \multicolumn{1}{c}{0.88} & 0.89  \\ \bottomrule
\end{tabular}
}}
\end{table}

\BfPara{Interpretation Models} 
Our study uses three interpreters: SHAP \cite{lundberg2017unified}, Saliency Maps \cite{simonyan2013deep}, and LIME \cite{ribeiro2016should}. Each interpreter has a unique methodology and provides valuable insights into model behavior and feature importance.

The use of different interpreters such as SHAP, Saliency Maps, and LIME enables a more comprehensive and deeper analysis of model performance and vulnerability to attacks, as each provides a unique perspective:

\begin{itemize}[leftmargin=1em]
    \item \textbf{SHAP}: By assigning consistent and fairly distributed importance values to each feature, SHAP helps identify which features are most influential in making predictions. This can reveal how the attack manipulates these key features to degrade model performance.

    \item \textbf{Saliency Maps}: By visualizing the gradient of the output with respect to the input features, Saliency Maps highlight which parts of the input data are most affected by the attack. This direct visualization allows us to see how the attack modifies the input to influence the model's decisions, providing insight into the attack's impact on critical features.

    \item {\textbf{LIME}: By creating interpretable surrogate models for specific predictions, LIME helps understand how the attack affects local decision boundaries. This localized view shows how the attack changes the model’s behavior in specific regions of the input space, highlighting potential weaknesses and areas where the model is  vulnerable.}
    
\end{itemize}

\subsection{Evaluation Metrics} 
We evaluated the attack success against DNN classifiers and interpretation models using multiple metrics:

\begin{itemize}[leftmargin=1em]
    \item \textbf{Attack Success Rate} This metric measures the effectiveness of the adversarial attack. It is calculated as the percentage of instances where the adversarial input successfully deceived the model, leading it to produce an incorrect output. A higher attack success rate indicates a more effective attack.
    \item \textbf{Number of Queries (\#)} This metric represents the number of queries made to the target model during the attack. It indicates the efficiency of the attack. Fewer queries suggest that the attack can deceive the model with minimal interaction. This is especially important in scenarios where the number of queries are limited or costly.
    \item \textbf{Misclassification Confidence} When the adversarial input is provided, this metric measures the confidence level of the model's incorrect prediction. A higher misclassification confidence indicates that the model is not only fooled by the adversarial input but is also highly sure about its wrong decision. This can provide insights into the model's vulnerability to the attack.
    \item \textbf{IoU (Intersection over Union)} Originally a metric used in computer vision tasks, IoU measures the overlap between two samples. In the context of NLP adversarial attacks, it can be used to quantify the similarity of interpretations between the original input and the adversarial input. A higher IoU indicates that the adversarial input retains much of the benign input's interpretation structure, suggesting a subtle and potentially more deceptive attack.
    \item \textbf{Avg. adversarial characters} This metric calculates the average number of characters in the input altered to craft the adversarial sample. It provides insights into the magnitude of changes required to deceive the model. A lower average suggests that minor perturbations are sufficient to mislead the model, indicating potential vulnerabilities in the model's understanding of the input.
    \item \textbf{Qualitative Metric (Manual Interpretation Comparison)} A qualitative assessment is crucial to understanding the real-world implications of the adversarial attack. For this metric, we manually compare the interpretations of samples in both their adversarial and benign forms. This involves assessing how the model's understanding or interpretation of the input changes due to the adversarial perturbations. Comparing the interpretations helps us understand the effectiveness of the attack, how much of the original meaning is kept, and how much the model's decision-making is affected. This hands-on approach gives a deeper insight into the attack's impact, showing where the model might be weak and providing a different perspective than just the numbers.
\end{itemize}

\section{Experiments and Evaluation} \label{sec:evaluation}
\BfPara{Dataset} For our experiment, we use three different datasets: \ed{Binary} Stanford Sentiment Treebank (SST-2) \cite{socher-etal-2013-recursive}, AG News \cite{Zhang2015CharacterlevelCN} \ed{and Yahoo Answers \cite{Zhang2015CharacterlevelCN}}. The Stanford Sentiment Treebank offers a detailed look into the sentiment structure of language through its labeled parse trees. The dataset contains 11,855 movie review sentences. Notably, it covers over 215,000 unique phrases, each evaluated by three human annotators. It is commonly termed SST-2 or SST binary for binary sentiment classification. The AG dataset contains over a million news articles with four news topic categories. 

\ed{The Yahoo Answers dataset, comprising approximately 1.4 million user-generated questions labeled for 10-category topic classification, introduces a highly challenging environment with its diverse, informal content. We use its test set to validate the attack's effectiveness in complex, multi-class settings.}

\begin{table}[ht]
\caption{\ed{The attack results of different classification models coupled with three interpreters on the SST-2, AG News and Yahoo Answers datasets. Results of Textbugger are repeated for different interpreters, as the attack does not consider attacking interpreter.
ASR, MC, PA, and QC denote attack success rate, misclassification confidence, charcter perturbation amount, and query count, respectively.}}
\label{tab:classifier_result}
\centering
{\color{black}
\resizebox{0.99\linewidth}{!}{%
\begin{tabular}{|cccccccccc|}
\hline
\multicolumn{1}{|c}{\multirow{2}{*}{}}          & \multicolumn{1}{|c}{\multirow{2}{*}{\textbf{Models}}}   & \multicolumn{4}{|c|}{\textbf{\ours}} & \multicolumn{4}{c|}{\textbf{Textbugger}} 
\\ \cline{3-10} \multicolumn{1}{|c}{} & \multicolumn{1}{|c}{} &  \multicolumn{1}{|c|}{ \textbf{ASR}} & \multicolumn{1}{c|}{\textbf{CQ}} & \multicolumn{1}{c|}{\textbf{MC}} & \multicolumn{1}{c|}{\textbf{PA}} & \multicolumn{1}{c|}{\makecell{\textbf{ASR}}} & \multicolumn{1}{c|}{\makecell{\textbf{QC}}} & \multicolumn{1}{c|}{\makecell{\textbf{MC}}} & \makecell{\textbf{PA}} \\ \hline
\multicolumn{10}{|c|}{\textbf{SST-2}}                                                                                                                                                                                                                  \\ \hline
\multicolumn{1}{|c|}{\multirow{7}{*}{\rotatebox{90}{SHAP}}} & \multicolumn{1}{c|}{GPT-2}      & \multicolumn{1}{c|}{0.69}                & \multicolumn{1}{c|}{20.99}      & \multicolumn{1}{c|}{93.18}                        & \multicolumn{1}{c|}{1.91}  & \multicolumn{1}{c|}{0.60} & \multicolumn{1}{c|}{61.50} & \multicolumn{1}{c|}{91.21} &  14.50                     \\  
\multicolumn{1}{|c|}{}                      & \multicolumn{1}{c|}{BERT}       & \multicolumn{1}{c|}{0.75}                & \multicolumn{1}{c|}{20.66}      & \multicolumn{1}{c|}{87.99}                        & \multicolumn{1}{c|}{1.85}  & \multicolumn{1}{c|}{0.79} & \multicolumn{1}{c|}{32.99} & \multicolumn{1}{c|}{93.84} & 8.33                                                  \\
\hline
\end{tabular}
}}
\end{table}

\begin{figure}[h]
    \centering
    \captionsetup{justification=justified}
    \includegraphics[width=1\linewidth]{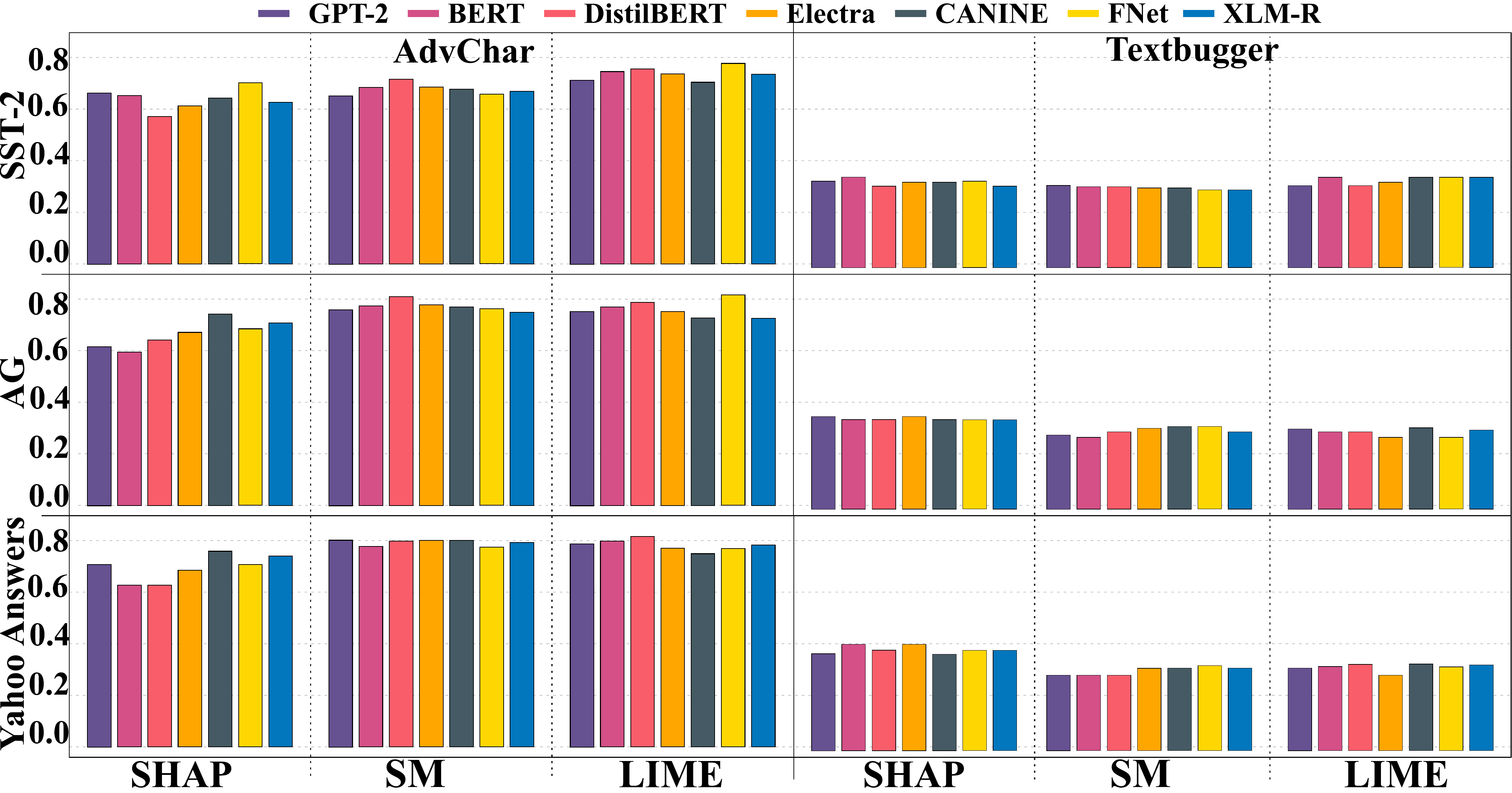}
    \caption{\ed{IoU scores of \ours{} and TextBugger~\cite{li2019textbugger} against seven classifiers with three interpreters on three datasets.}}\vspace{-1em}
    \label{fig:iou_score}
\end{figure}
\subsection{Attack Effectiveness against Classifiers}
\ed{\autoref{tab:classifier_result} presents the attack results across the seven NLP models on SST-2, AG News, and Yahoo Answers, highlighting its effectiveness with different interpreters. We also compare our results with the existing attack TextBugger {\cite{li2019textbugger}}. For this comparison, we use the default settings provided in TextBugger's original implementation {\cite{li2019textbugger}}. 
TextBugger is selected as a baseline because it is one of the most widely used and well-established character-level black-box attacks for text classification tasks.} For the SST-2 dataset, the SHAP interpreter has an attack success rate of approximately 64.71\%. It required, on average, 20.68 queries with misclassification confidence of around 89.91\% and an average number of 1.89 adversarial characters. The attack with SM interpreter performed better, with the highest attack success rate of 73.43\%. However, a considerable number of queries, averaging 60.93, were required. Its misclassification confidence and average adversarial characters were approximately 89.50\% and 2.14, respectively. While achieving an attack success rate of about 69.57\%, the LIME interpreter was more efficient than the others regarding the number of queries, averaging only 17.76. It has the highest misclassification confidence at around 90.38\% and the lowest average adversarial characters at 1.72.

For the AG News dataset, the SHAP interpreter's performance showed an attack success rate of 53.29\%, the lowest among the interpreters. It required an average of 40.75 queries and had a misclassification confidence of around 56.91\%. It also generated more adversarial characters for the dataset, averaging 3.65. With an attack success rate of roughly 69.00\%, the SM interpreter had a significantly high number of queries, averaging 258.84, but had misclassification confidence of around 58.51\% and average adversarial characters of 2.13. The LIME interpreter was the most effective for this dataset, resulting in the highest attack success rate of 74.57\%. It was also efficient, requiring the fewest queries, with an average of 34.45. Its misclassification confidence was around 81.78\%, with the lowest average adversarial characters at 1.80.

Regarding the SST-2 dataset classifiers, BERT showed superior performance, with a 75.00\% success rate and an average of 32.91 queries per attack. The classifier's misclassification confidence was at 88.22\%, with an average of 1.87 adversarial characters. With an attack success rate of 71.33\%, CANINE demonstrated efficiency by averaging 31.55 queries. Interestingly, the attack required the least number of adversarial character perturbations, averaging 1.73 against this model with high misclassification confidence of 93.27\%. Contrarily, DistilBERT was found to be more robust regarding misclassification, achieving a success rate of 69.33\% with a misclassification confidence of 79.75\%. Electra has a success rate of 68.00\% and higher query usage of an average of 34.63. FNet and GPT-2 classifiers had 66.00\% and 68.33\% attack success rates, respectively. Meanwhile, XLM-R had an average attack success rate of 66.67\% but was notable for its high misclassification confidence, reaching 96.30\%.

For the classifiers applied to the AG News dataset, their behavior demonstrated significant changes. While maintaining an attack success rate of 67.33\%, BERT required an increased 110.71 queries on average, showing a dataset-specific behavior. Its misclassification confidence was 56.14\%. CANINE, reporting an attack success rate of 68.67\%, performed with misclassification confidence of 72.38\%, while DistilBERT provided the lowest misclassification confidence of 53.94\% despite its success rate of 66.33\%. Electra's attack success rate was 65.67\%, but its query requirement was high at an average of 116.63. The FNet model presented a success rate of 63.33\%, while GPT-2 displayed a 61.00\% success rate with a misclassification confidence of 62.32\%. XLM-R achieved a 67.00\% success rate while maintaining a 72.52\% misclassification confidence.

\ed{Extending the analysis to the Yahoo Answers dataset, we observe that the attack's performance varies across interpreters and models. The SHAP interpreter exhibits lower attack success rates, ranging from 45\% to 55\%, with higher query requirements (approximately 42 to 50) and more adversarial characters (around 3.63 to 4.17) compared to its performance on SST-2. This could be attributed to the increased complexity and length of the dataset inputs, which may pose greater challenges for generating effective adversarial examples. The SM interpreter achieves success rates between 60\% and 71\%, but at the cost of a substantial increase in the number of queries, averaging around 280 to 311, indicating a trade-off between success and efficiency. The LIME interpreter achieves the highest success rates, from 75\% to 81\%, while maintaining relatively low query counts (approximately 36 to 43) and minimal adversarial perturbations (around 1.97 to 2.31), suggesting it adapts particularly well to this dataset.}

\ed{Our method consistently outperforms TextBugger across the datasets of SST-2, AG News, and Yahoo Answers. On SST-2, it achieves comparable or higher success rates (\eg 0.79 vs. 0.79 for BERT with SM), while requiring significantly fewer queries (\eg 17.18 vs. 34.54 for CANINE with LIME) and fewer adversarial character modifications (\eg 1.63 vs. 8.46), along with higher misclassification confidence (\eg 97.11\% vs. 94.36\% for XLM-R with LIME). For AG News, \ours  demonstrates a higher success rate (\eg 0.79 vs. 0.58 for CANINE with LIME) with fewer queries (\eg 33.31 vs. 94.30) and minimal perturbations (\eg 1.73 vs. 46.70). However, SM occasionally requires more queries (\eg 256.23 vs. 102.92 for BERT). On Yahoo Answers, our method achieves superior success (\eg 0.81 vs. 0.49 for CANINE with LIME), lower query counts (\eg 36.77 vs. 156.44), and reduced perturbations (\eg 1.98 vs. 88.72), while also producing higher confidence scores (\eg 94.90\% vs. 58.10\%). Notably, the LIME interpreter contributes significantly to this efficiency across datasets, while SHAP and SM offer robust success rates despite occasional increases in query count, highlighting \ours's adaptability and efficiency over TextBugger.}

\ed{\ours's superior performance over TextBugger can be attributed to three key factors. First, unlike TextBugger, which uses visual character substitutions (\eg replacing ``o'' with ``0''), our method leverages interpreter-driven importance scores from SHAP, SM, and LIME to target semantically critical tokens, ensuring minimal character changes significantly alter model predictions. Second, modern NLP models, pretrained on noisy web-scale data, can be robust to TextBugger's visual perturbations, reducing its effectiveness. 
Third, by avoiding random perturbations and preserving interpreter consistency, our method maintains stealthiness and efficiency, outperforming TextBugger in success rate, query count, and perturbation minimization across diverse models.}

\begin{figure*}[h]
    \centering
    \captionsetup{justification=justified}
    \includegraphics[width=\linewidth]{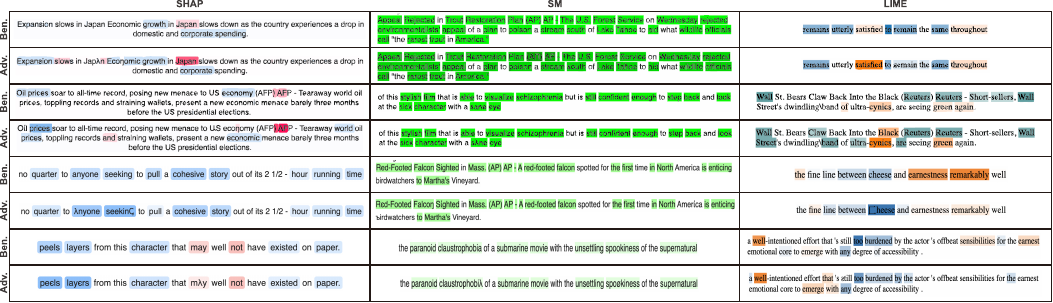}
    \caption{\ed{Results of interpreters on benign and adversarial texts generated by \ours against BERT, GPT and Electra models. Ben. and Adv. stand for benign and adversarial texts, respectively. 
    Samples are selected randomly from SST-2 and AG datasets.}}\vspace{-1em}
    \label{fig:example_text}
\end{figure*}

\begin{figure}[h!]
 \begin{subfigure}{0.45\textwidth}
  \centering
  \includegraphics[width=0.95\textwidth]{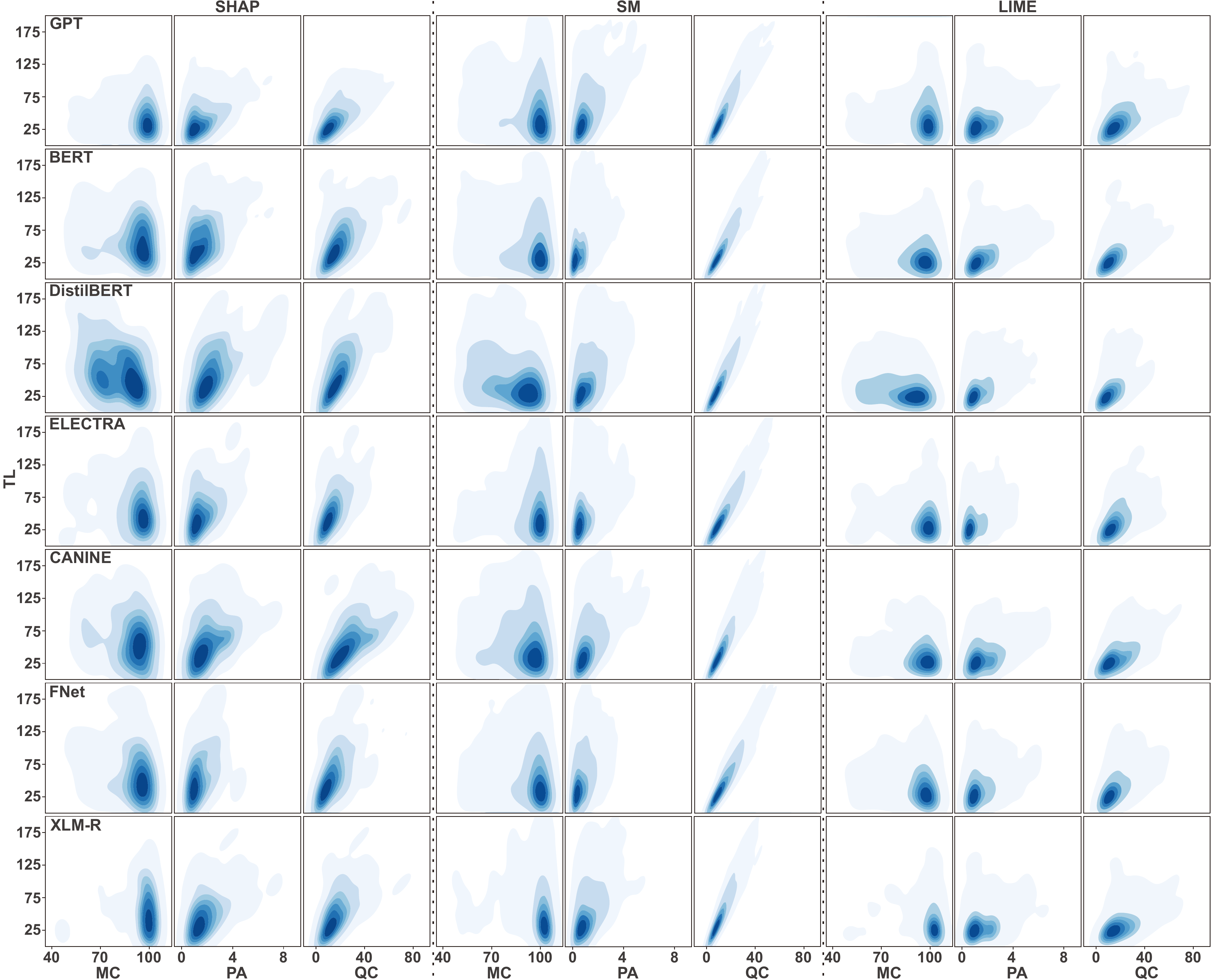}
  \subcaption{SST-2 dataset.}
 \label{fig:lt_mc_qc}
 \end{subfigure} 
 \begin{subfigure}{0.45\textwidth}
     \centering
    \includegraphics[width=0.95\textwidth]{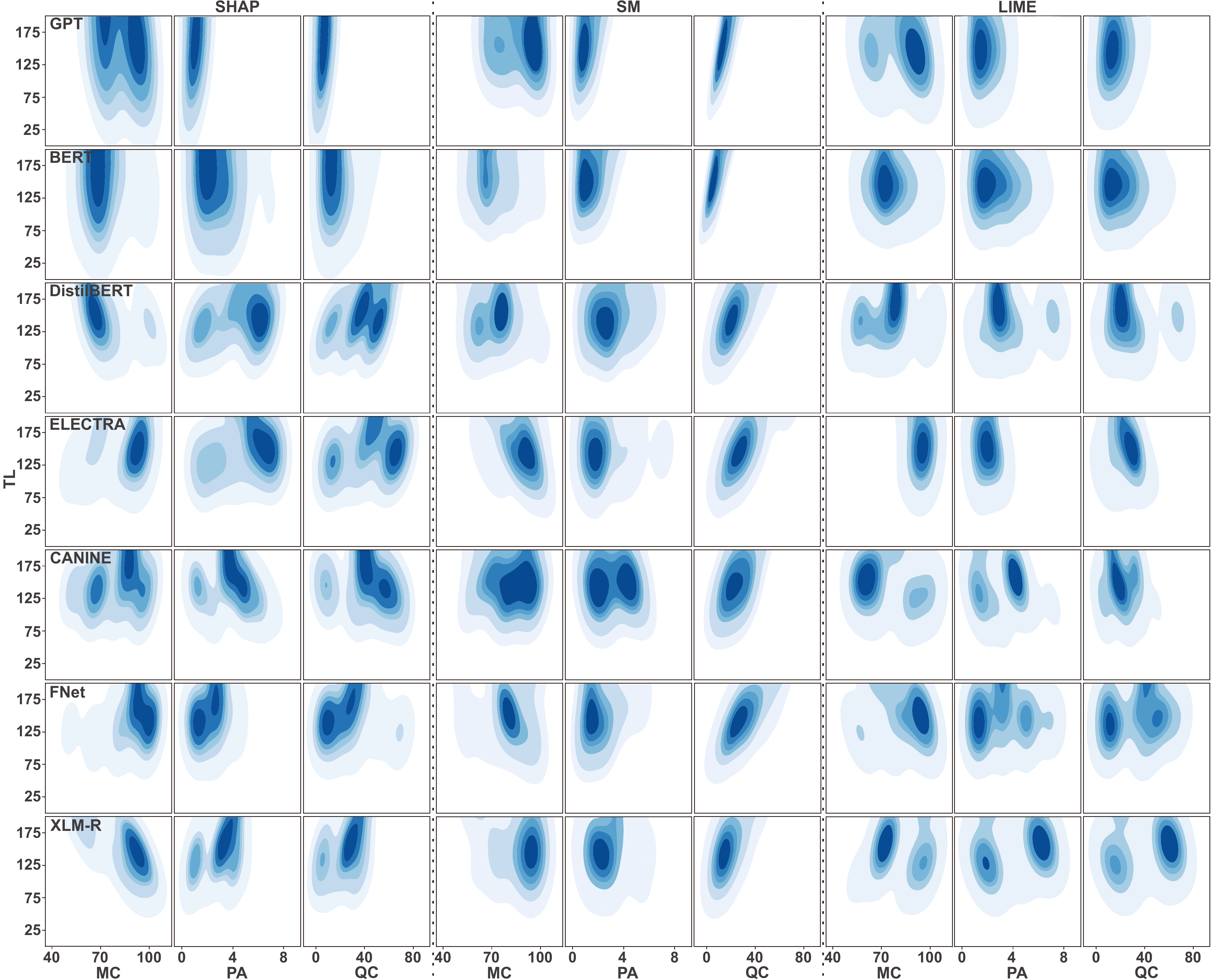}
    \subcaption{AG News dataset.}
    \label{fig:lt_mc_qc_news}
    \end{subfigure}
\begin{subfigure}{0.45\textwidth}
     \centering
    \includegraphics[width=0.95\textwidth]{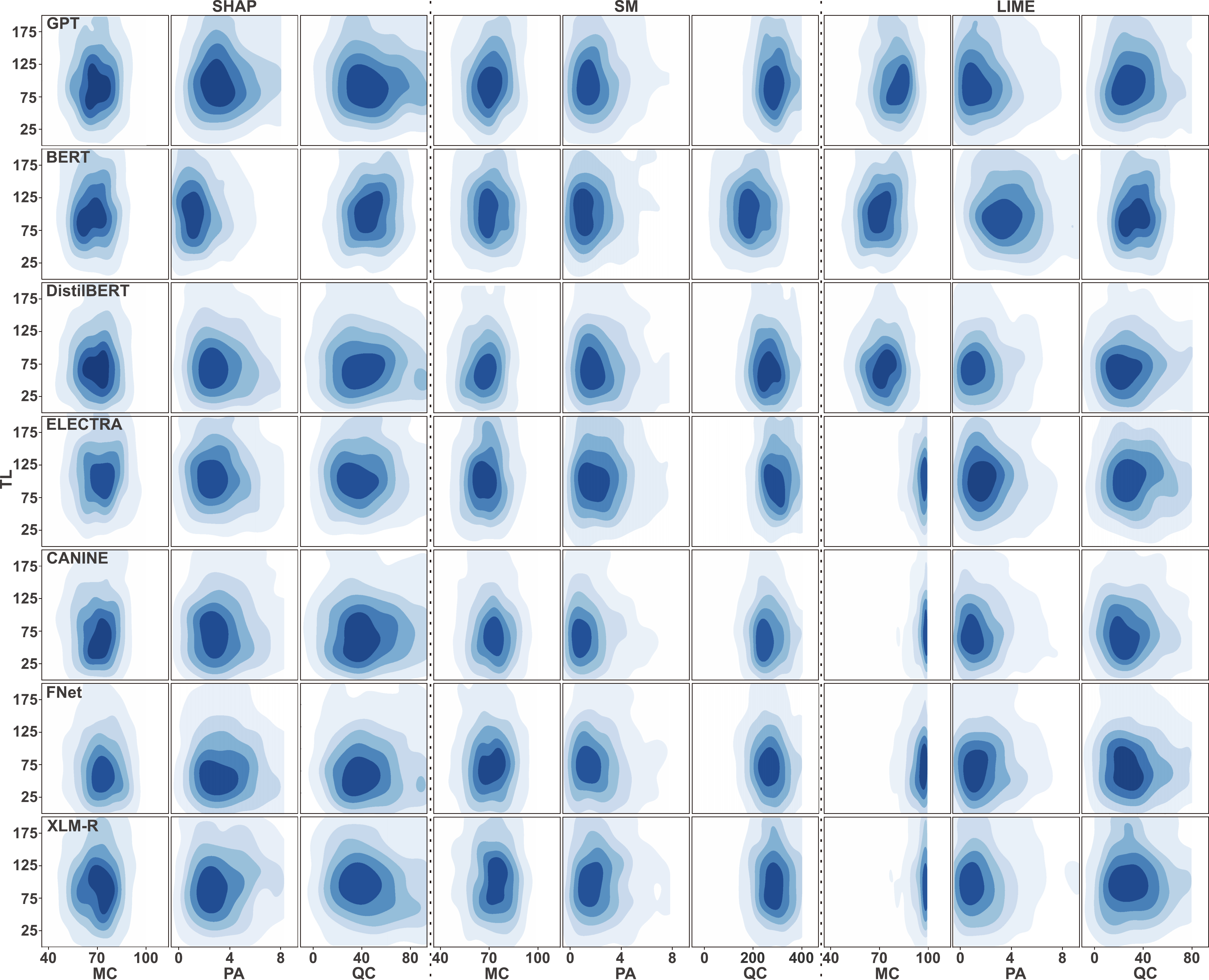}
    \subcaption{\ed{Yahoo Answers dataset.}}
    \label{fig:lt_mc_qc_yahoo}
    \end{subfigure}
    \caption{Density plots showcasing the distribution of input text length (TL) concerning misclassification confidence, perturbation amount, and query count (MC, PA, and QC) for different models, using three interpreters - SHAP, SM, and LIME.}
\end{figure} 

\subsection{Attack Effectiveness against Interpreters}
Our analysis employed the Intersection over Union (IoU) scores to measure the similarity between adversarial and benign interpretations across various NLP models. Seven models were assessed on the SST-2 and AG News datasets using three interpreters: SHAP, SM, and LIME. \autoref{fig:iou_score} displays the results of IoU scores in our experiment.

On the SST-2 dataset with the SHAP interpreter, the FNet model registered the peak IoU score at 0.70, with GPT-2 and BERT closely following at 0.66 and 0.65, respectively. DistilBERT indicated adversarial interpretation with the lowest score of 0.57. Under the SM interpreter, DistilBERT achieved the highest IoU score of 0.71, while models like GPT-2, BERT, and Electra produced scores ranging from 0.65 to 0.68. When using the LIME interpreter, FNet scored the highest, 0.77, while GPT-2 and BERT scored above 0.70.

For the AG News dataset, CANINE  with the SHAP interpreter performed with an IoU score of 0.72, contrasting BERT at the lower range with 0.59. The top performers among SM interpreters were DistilBERT, with a score of 0.81, followed by BERT and Electra, with scores of 0.78 each. FNet and DistilBERT scored 0.81 and 0.79 with the LIME, respectively.

\ed{For the Yahoo Answers dataset, \ours continues to demonstrate high interpretation similarity across all models and interpreters. Under SHAP, IoU scores range from 0.61 to 0.69, with FNet and XLM-R achieving the highest similarity. Using SM, most models produce IoU scores between 0.75 and 0.81. The LIME interpreter again shows the highest IoU values, with several models including BERT, GPT-2, and Electra reaching or exceeding 0.80.}

\ed{In comparison with TextBugger~\cite{li2019textbugger} using the same experimental setup, \autoref{fig:iou_score} shows that TextBugger produces lower IoU scores across all datasets. 
For example, on SST-2, TextBugger's IoU scores range between 0.28 to 0.36, indicating limited similarity between adversarial and benign interpretations. This contrast is especially noticeable under the LIME interpreter, where our method consistently exceeds 0.75, while TextBugger remains below 0.36 for all models. Similar results are observed on AG News and Yahoo Answers datasets.} 

We also evaluate the capability of our attack to create interpretation maps comparable to benign interpretation maps. We conduct a qualitative evaluation by manual checking to observe the similarity of interpretations produced by adversarial and the corresponding benign samples. By qualitatively comparing the interpretations of benign and adversarial samples, our attack-generated interpretations are visually highly similar to their corresponding benign sample inputs. \autoref{fig:example_text} shows examples of observed attribution maps obtained using SHAP, SM, and LIME. The examples show adversarial interpretations and corresponding benign ones. The adversarial and benign attribution maps are highly similar as displayed in \autoref{fig:example_text}.

\begin{figure*}[th!]
 \begin{minipage}{0.33\textwidth}
  \centering
  \includegraphics[width=1\textwidth]{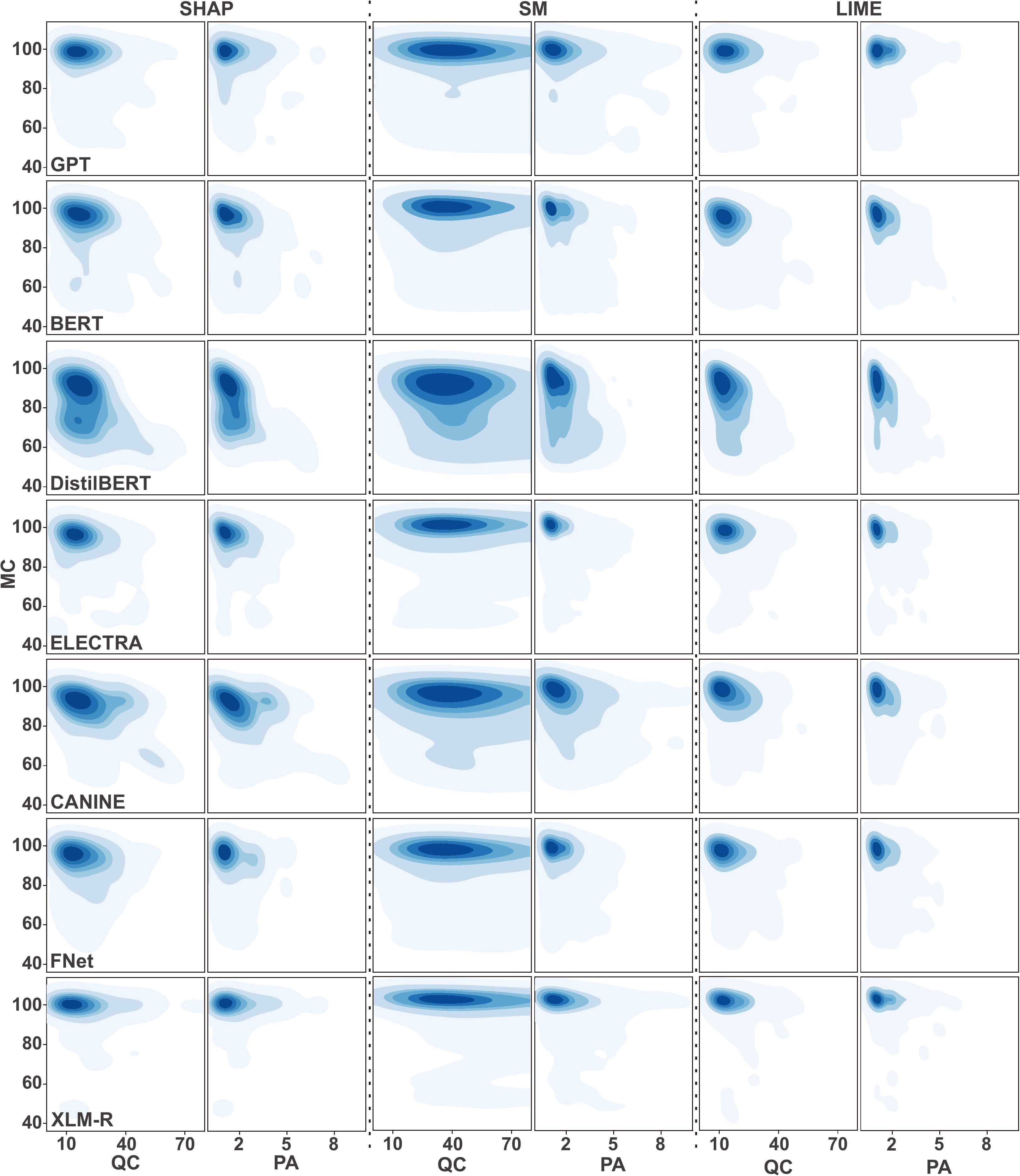}
  \subcaption{SST-2 dataset.}
 \label{fig:mc_qc_pa}
 \end{minipage}~
 \begin{minipage}{0.33\textwidth}
     \centering
    \captionsetup{justification=justified}
    \includegraphics[width=1\textwidth]{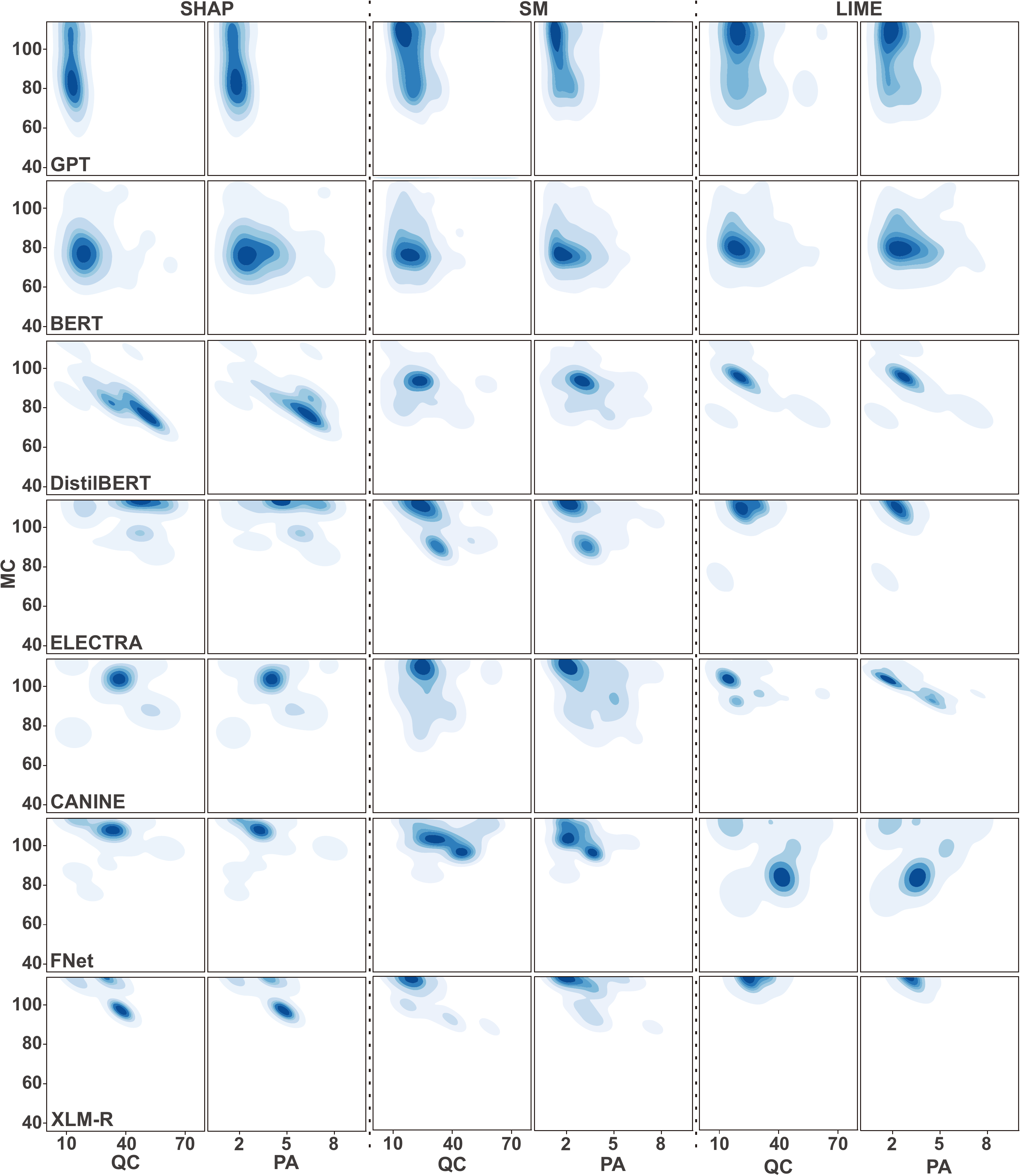}
    \subcaption{AG News dataset.}
    \label{fig:mc_qc_pa_news}
    \end{minipage}~
\begin{minipage}{0.33\textwidth}
     \centering
    \captionsetup{justification=justified}
    \includegraphics[width=1\textwidth]{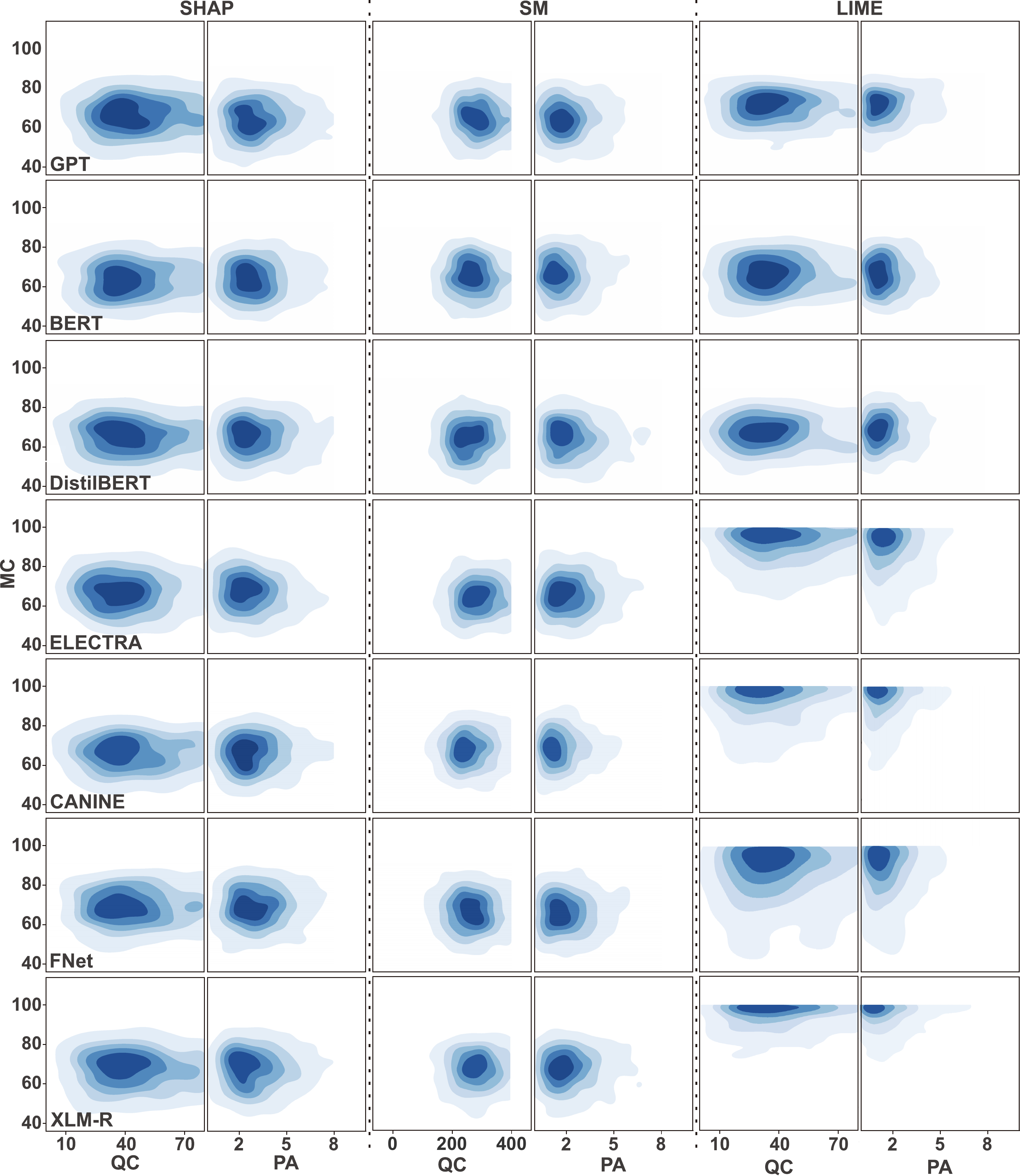}
    \subcaption{\ed{Yahoo Answers dataset.}}
    \label{fig:mc_qc_pa_yahoo}
    \end{minipage}
 \caption{\ed{Density plots showcasing the distribution of misclassification confidence (MC) concerning query count and perturbation amount (QC and PA) for different models, using three interpreters - SHAP, SM, and LIME.}}
\end{figure*}

\subsection{Attack Transferability}
Adversarial inputs have a desirable property of transferability. This means that an adversarial input effective against one deep neural network (DNN) can also be effective against other DNNs. Our study aims to explore the transferability of our attacks across interpreters and classifiers. 

\begin{table}[]
\caption{\ed{IoU scores of the transferability of adversarial inputs from source interpreters to target interpreters on different SST-2, AG News, and Yahoo Answers datasets classifiers.}}
\label{tab:int_transferability}
\centering
{\color{black}
\resizebox{\linewidth}{!}{%
\begin{tabular}{|ccccccccc|}
\hline
\multicolumn{1}{|c|}{\multirow{2}{*}{\textbf{\begin{tabular}[c]{@{}c@{}}Source \end{tabular}}}} & \multicolumn{1}{c|}{\multirow{2}{*}{\textbf{\begin{tabular}[c]{@{}c@{}}Target\end{tabular}}}} & \multicolumn{7}{c|}{\textbf{Classifiers}} \\ \cline{3-9} 
\multicolumn{1}{|c|}{} & \multicolumn{1}{c|}{} & \multicolumn{1}{c|}{\textbf{GPT-2}} & \multicolumn{1}{c|}{\textbf{BERT}} & \multicolumn{1}{c|}{\textbf{DistilBERT}} & \multicolumn{1}{c|}{\textbf{Electra}} & \multicolumn{1}{c|}{\textbf{CANINE}} & \multicolumn{1}{c|}{\textbf{FNet}} & \textbf{XLM-R} \\ \hline
\multicolumn{9}{|c|}{\textbf{SST-2}} \\ \hline
\multicolumn{1}{|c|}{\multirow{2}{*}{\textbf{SHAP}}} & \multicolumn{1}{c|}{\textbf{LIME}} & \multicolumn{1}{c|}{0.23} & \multicolumn{1}{c|}{0.30} & \multicolumn{1}{c|}{0.15} & \multicolumn{1}{c|}{0.13} & \multicolumn{1}{c|}{0.15} & \multicolumn{1}{c|}{0.16} & 0.15 \\   
\multicolumn{1}{|c|}{} & \multicolumn{1}{c|}{\textbf{SM}} & \multicolumn{1}{c|}{0.21} & \multicolumn{1}{c|}{0.27} & \multicolumn{1}{c|}{0.13} & \multicolumn{1}{c|}{0.12} & \multicolumn{1}{c|}{0.14} & \multicolumn{1}{c|}{0.13} & 0.13 \\ \hline
\multicolumn{1}{|c|}{\multirow{2}{*}{\textbf{SM}}} & \multicolumn{1}{c|}{\textbf{SHAP}} & \multicolumn{1}{c|}{0.19} & \multicolumn{1}{c|}{0.34} & \multicolumn{1}{c|}{0.23} & \multicolumn{1}{c|}{0.26} & \multicolumn{1}{c|}{0.22} & \multicolumn{1}{c|}{0.20} & 0.27 \\   
\multicolumn{1}{|c|}{} & \multicolumn{1}{c|}{\textbf{LIME}} & \multicolumn{1}{c|}{0.23} & \multicolumn{1}{c|}{0.62} & \multicolumn{1}{c|}{0.31} & \multicolumn{1}{c|}{0.24} & \multicolumn{1}{c|}{0.23} & \multicolumn{1}{c|}{0.21} & 0.28 \\ \hline
\multicolumn{1}{|c|}{\multirow{2}{*}{\textbf{LIME}}} & \multicolumn{1}{c|}{\textbf{SHAP}} & \multicolumn{1}{c|}{0.12} & \multicolumn{1}{c|}{0.12} & \multicolumn{1}{c|}{0.14} & \multicolumn{1}{c|}{0.11} & \multicolumn{1}{c|}{0.13} & \multicolumn{1}{c|}{0.14} & 0.12 \\   
\multicolumn{1}{|c|}{} & \multicolumn{1}{c|}{\textbf{SM}} & \multicolumn{1}{c|}{0.17} & \multicolumn{1}{c|}{0.15} & \multicolumn{1}{c|}{0.18} & \multicolumn{1}{c|}{0.14} & \multicolumn{1}{c|}{0.16} & \multicolumn{1}{c|}{0.16} & 0.16 \\ \hline
\multicolumn{9}{|c|}{\textbf{AG}} \\ \hline
\multicolumn{1}{|c|}{\multirow{2}{*}{\textbf{SHAP}}} & \multicolumn{1}{c|}{\textbf{LIME}} & \multicolumn{1}{c|}{0.31} & \multicolumn{1}{c|}{0.57} & \multicolumn{1}{c|}{0.24} & \multicolumn{1}{c|}{0.23} & \multicolumn{1}{c|}{0.27} & \multicolumn{1}{c|}{0.30} & 0.27 \\   
\multicolumn{1}{|c|}{} & \multicolumn{1}{c|}{\textbf{SM}} & \multicolumn{1}{c|}{0.22} & \multicolumn{1}{c|}{0.51} & \multicolumn{1}{c|}{0.21} & \multicolumn{1}{c|}{0.20} & \multicolumn{1}{c|}{0.20} & \multicolumn{1}{c|}{0.21} & 0.21 \\ \hline
\multicolumn{1}{|c|}{\multirow{2}{*}{\textbf{SM}}} & \multicolumn{1}{c|}{\textbf{SHAP}} & \multicolumn{1}{c|}{0.34} & \multicolumn{1}{c|}{0.52} & \multicolumn{1}{c|}{0.39} & \multicolumn{1}{c|}{0.34} & \multicolumn{1}{c|}{0.35} & \multicolumn{1}{c|}{0.34} & 0.46 \\   
\multicolumn{1}{|c|}{} & \multicolumn{1}{c|}{\textbf{LIME}} & \multicolumn{1}{c|}{0.35} & \multicolumn{1}{c|}{0.65} & \multicolumn{1}{c|}{0.56} & \multicolumn{1}{c|}{0.46} & \multicolumn{1}{c|}{0.30} & \multicolumn{1}{c|}{0.38} & 0.53 \\ \hline
\multicolumn{1}{|c|}{\multirow{2}{*}{\textbf{LIME}}} & \multicolumn{1}{c|}{\textbf{SHAP}} & \multicolumn{1}{c|}{0.24} & \multicolumn{1}{c|}{0.21} & \multicolumn{1}{c|}{0.27} & \multicolumn{1}{c|}{0.26} & \multicolumn{1}{c|}{0.27} & \multicolumn{1}{c|}{0.25} & 0.26 \\   
\multicolumn{1}{|c|}{} & \multicolumn{1}{c|}{\textbf{SM}} & \multicolumn{1}{c|}{0.32} & \multicolumn{1}{c|}{0.29} & \multicolumn{1}{c|}{0.32} & \multicolumn{1}{c|}{0.22} & \multicolumn{1}{c|}{0.21} & \multicolumn{1}{c|}{0.22} & 0.29 \\ \hline
\multicolumn{9}{|c|}{\textbf{Yahoo Answers}} \\ \hline
\multicolumn{1}{|c|}{\multirow{2}{*}{\textbf{SHAP}}} & \multicolumn{1}{c|}{\textbf{LIME}} & \multicolumn{1}{c|}{0.35} & \multicolumn{1}{c|}{0.62} & \multicolumn{1}{c|}{0.28} & \multicolumn{1}{c|}{0.27} & \multicolumn{1}{c|}{0.31} & \multicolumn{1}{c|}{0.34} & 0.31 \\   
\multicolumn{1}{|c|}{} & \multicolumn{1}{c|}{\textbf{SM}} & \multicolumn{1}{c|}{0.25} & \multicolumn{1}{c|}{0.58} & \multicolumn{1}{c|}{0.26} & \multicolumn{1}{c|}{0.25} & \multicolumn{1}{c|}{0.24} & \multicolumn{1}{c|}{0.24} & 0.24 \\ \hline
\multicolumn{1}{|c|}{\multirow{2}{*}{\textbf{SM}}} & \multicolumn{1}{c|}{\textbf{SHAP}} & \multicolumn{1}{c|}{0.38} & \multicolumn{1}{c|}{0.56} & \multicolumn{1}{c|}{0.44} & \multicolumn{1}{c|}{0.39} & \multicolumn{1}{c|}{0.39} & \multicolumn{1}{c|}{0.39} & 0.50 \\   
\multicolumn{1}{|c|}{} & \multicolumn{1}{c|}{\textbf{LIME}} & \multicolumn{1}{c|}{0.40} & \multicolumn{1}{c|}{0.71} & \multicolumn{1}{c|}{0.61} & \multicolumn{1}{c|}{0.51} & \multicolumn{1}{c|}{0.34} & \multicolumn{1}{c|}{0.42} & 0.59 \\ \hline
\multicolumn{1}{|c|}{\multirow{2}{*}{\textbf{LIME}}} & \multicolumn{1}{c|}{\textbf{SHAP}} & \multicolumn{1}{c|}{0.27} & \multicolumn{1}{c|}{0.24} & \multicolumn{1}{c|}{0.31} & \multicolumn{1}{c|}{0.29} & \multicolumn{1}{c|}{0.30} & \multicolumn{1}{c|}{0.26} & 0.29 \\   
\multicolumn{1}{|c|}{} & \multicolumn{1}{c|}{\textbf{SM}} & \multicolumn{1}{c|}{0.36} & \multicolumn{1}{c|}{0.33} & \multicolumn{1}{c|}{0.36} & \multicolumn{1}{c|}{0.28} & \multicolumn{1}{c|}{0.25} & \multicolumn{1}{c|}{0.25} & 0.32 \\ \hline
\end{tabular}
}}
\end{table}

\begin{table*}[]
\caption{\ed{Transferability of adversarial inputs across NLP classifiers, measured by Attack Success Rate (ASR) and Misclassification Confidence (MC) on SST-2, AG News, and Yahoo Answers datasets. The first and second columns represent the source classifiers with their coupled interpreters, while the first row shows the target classifiers.}}
\label{tab:cls_transferability}
\centering
{\color{black}
\resizebox{0.65\linewidth}{!}{%
\begin{tabular}{|cccccccccccccccc|}
\hline
\multicolumn{1}{|c|}{}                               & \multicolumn{1}{c|}{}                    & \multicolumn{2}{c|}{\textbf{GPT-2}}                                  & \multicolumn{2}{c|}{\textbf{BERT}}                                   & \multicolumn{2}{c|}{\textbf{DistilBERT}}                             & \multicolumn{2}{c|}{\textbf{Electra}}                                & \multicolumn{2}{c|}{\textbf{CANINE}}                                 & \multicolumn{2}{c|}{\textbf{FNet}}                                   & \multicolumn{2}{c|}{\textbf{XLM-R}}             \\ \hline
\multicolumn{1}{|c|}{}                               & \multicolumn{1}{c|}{}                    & \multicolumn{1}{c|}{\textbf{ASR}} & \multicolumn{1}{c|}{\textbf{MC}} & \multicolumn{1}{c|}{\textbf{ASR}} & \multicolumn{1}{c|}{\textbf{MC}} & \multicolumn{1}{c|}{\textbf{ASR}} & \multicolumn{1}{c|}{\textbf{MC}} & \multicolumn{1}{c|}{\textbf{ASR}} & \multicolumn{1}{c|}{\textbf{MC}} & \multicolumn{1}{c|}{\textbf{ASR}} & \multicolumn{1}{c|}{\textbf{MC}} & \multicolumn{1}{c|}{\textbf{ASR}} & \multicolumn{1}{c|}{\textbf{MC}} & \multicolumn{1}{c|}{\textbf{ASR}} & \textbf{MC} \\ \hline
\multicolumn{16}{|c|}{\textbf{SST-2}}                                                                                                                                                                                                                                                                                                                                                                                                                                                                                                                                                         \\
\hline
\end{tabular}
}}
\end{table*}

\BfPara{Interpreter} We generate a set of adversarial inputs against the source interpreter, denoted by $G$, and use them to calculate their attribution maps using target interpreters, denoted by $G'$. In exploring the transferability of adversarial inputs using the SST-2 and AG News datasets, distinct patterns emerged. The IoU scores quantitatively measured how similar adversarial examples were recognized between different interpreters. 

A detailed overview of the IoU scores across different interpreters is provided in \autoref{tab:int_transferability}. As shown in the table, BERT demonstrated higher similarity in its attribution maps with an average IoU score of  0.3792. This suggests that adversarial examples generated for BERT tend to keep similarity by other interpreters. However, CANINE showed more varied interpretability with an average IoU score of  0.2192, indicating reduced transferability of its adversarial examples across interpreters. GPT-2 also showed low transferability in specific scenarios, \eg adversarial examples from LIME measured by SHAP, resulting in a low IoU score of 0.12. 

When we explore the usage of specific interpreter combinations., the SM-LIME pairing frequently indicated high transferability, particularly in the SST-2 dataset, where it achieved IoU scores as high as  0.62 for BERT. However, adversarial examples from LIME, when measured by SHAP, often showed lower transferability, with IoU scores dropping to 0.12 for BERT and GPT-2 in the SST-2 dataset. 

For the AG News dataset, both SM-LIME and SM-SHAP combinations showcased strong transferability, with IoU scores reaching 0.65 and 0.52 for BERT, respectively. The difference between the datasets can be the result of their unique characteristics. SST-2 focuses on sentiment analysis, while AG News involves news article categorization. Such differences can influence the generation and interpretability of adversarial examples.

\ed{For the Yahoo Answers dataset, transferability patterns align with those observed in AG News, with BERT showing high similarity in attribution maps, achieving IoU scores up to 0.71 for SM to LIME and 0.62 for SHAP to LIME. The SM-LIME pairing remains the most transferable, with scores like 0.61 for DistilBERT and 0.59 for XLM-R, while LIME to SHAP produces low transferability, with scores as low as 0.24 for BERT and 0.27 for GPT-2. CANINE continues to exhibit varied interpretability, with IoU scores ranging from 0.24 (SHAP to SM) to 0.39 (SM to SHAP), and GPT-2 shows moderate transferability, reaching 0.40 for SM to LIME.}

\BfPara{Classifier} After examining the transferability of adversarial inputs across various NLP classifiers using the SST-2 and AG News datasets, \autoref{tab:cls_transferability} provides insights into the effectiveness of adversarial attacks and the confidence with which classifiers make misclassifications. Two metrics quantify this: Attack Success Rate (ASR) and Misclassification Confidence (MC). 

The SHAP interpreter paired with Electra as the source classifier showed high adversarial transferability, especially when targeting BERT. In the SST-2 dataset, this pairing achieved an ASR of 0.72 against BERT, while in the AG News dataset, an ASR of 0.66 was recorded against DistilBERT.

The SST-2 dataset highlighted SHAP with CANINE targeting GPT-2 and XLM-R as the least transferable combination with an ASR of 0.1. In contrast, the SM interpreter with CANINE targeting Electra in the AG News dataset achieved a minimal ASR of 0.04.
\ed{For the Yahoo Answers dataset, SHAP with Electra performs strong transferability, achieving an ASR of 0.50 against BERT (MC 0.45), consistent with SST-2 and AG News trends. LIME with DistilBERT also shows high transferability, with an ASR of 0.55 against BERT (MC 0.42). In contrast, SHAP and SM with CANINE are the least transferable, with ASRs as low as 0.11–0.12 (MC 0.09–0.10). }
The results show the significant adversarial transferability potential of certain combinations of models, with BERT often emerging as a susceptible target. 

\subsection{Comparative Metric Analysis}
Identifying trends can offer critical insights into optimal model selection and tuning when examining different architectures and their reactions to various input conditions. We embarked on a comparative analysis to shed light on this, focusing primarily on model architectures such as GPT, BERT, DistilBERT, ELECTRA, CANINE, FNet, and XLM-R. We calculated the correlation between input text length (TL), misclassification confidence (MC), query count (QC), and perturbation amount (PA) through a detailed assessment. Figures~\ref{fig:lt_mc_qc}--\ref{fig:mc_qc_pa_news} present a visual representation of our findings, highlighting the specific behaviors of each model across different interpreters (SHAP, SM, and LIME) on two datasets.

In \autoref{fig:lt_mc_qc}, when analyzed using SHAP on the SST-2 dataset, most models display a concentrated cluster of values in the mid- to high-range of MC scores, regardless of the length of the input text. Notably, the density plots exhibit a vertical pattern, which suggests that the MC score is usually consistent, irrespective of the input text's length. The behavior of models analyzed using SM is more varied compared to SHAP. Density plots show horizontal and vertical orientations, indicating that the interpretation heavily depends on the model and its interaction with the input text length. LIME interpretations reveal a combination of patterns found in SHAP and SM. Some models show dense concentration around specific MC values (like SHAP), while others exhibit variations across text lengths (like SM). When analyzed using SHAP and LIME, models like GPT and BERT show a high concentration around mid-range MC values, regardless of input text length. This suggests a consistent confidence level in misclassification maintained across different text lengths. The behavior of the DistilBERT and ELECTRA models seems to be influenced by the interpretation method. At the same time, SHAP interpretations result in consistent MC values, and SM and LIME exhibit more varied patterns, particularly about text length. CANINE, FNet, and XLM-R models show diverse performances, especially under SM. Density plots indicate MC and input text length influence interpretation. In \autoref{fig:lt_mc_qc_news}, When analyzed using SHAP on the AG News dataset, the models demonstrate distinct vertical patterns, particularly GPT, BERT, and XLM-R. This indicates a consistency in MC regardless of the input length. For DistilBERT and ELECTRA, there is a noticeable split in the density, which implies two potential clusters or behaviors based on MC, leading to non-uniform model responses across data points. The density patterns of SM vary more than SHAP. GPT and BERT show diverse patterns based on text length, while ELECTRA and CANINE display dual peaks, indicating two potential behaviors. 
A combination of dense and dispersed areas can be observed when using LIME with models such as GPT and BERT. However, the pattern of CANINE is remarkably distinct. It has an oval shape with high density at mid-range MC values, which suggests a strong correlation between MC and a specific range of text lengths.

\ed{In \autoref{fig:lt_mc_qc_yahoo}, the SHAP-based analysis on the Yahoo Answers dataset reveals that models such as GPT-2, BERT, and XLM-R exhibit vertically aligned density patterns, with misclassification confidence (MC) values concentrated in the mid to high range (approximately 40–80), irrespective of input text length. This suggests stable confidence in misclassification across varying input sizes. In contrast, DistilBERT and ELECTRA show more dispersed MC distributions, indicating greater sensitivity to input length. Under the SM interpreter, density patterns become more diverse, for instance, GPT-2 and BERT demonstrate dual MC peaks around 20 and 70, while CANINE and FNet exhibit broader density regions associated with high query counts (QC), ranging from 200 to 400 with the increased computational cost. The LIME-based results also show a mix of behaviors: GPT-2 and BERT maintain consistent MC values in the 40–80 range, whereas CANINE forms a distinct oval-shaped cluster centered around MC values of 40–60 and text lengths between 75 and 125.}

In \autoref{fig:mc_qc_pa}, The MC distribution on the SST-2 dataset for all models in SHAP displays primarily elliptical patterns. For models like GPT and BERT, the graphs based on QC have a solid vertical alignment, suggesting consistent misclassification confidence for varied query counts.
Models with SM produce broader and more diverse patterns, whereas GPT and BERT still exhibit vertical concentrations. CANINE and FNet, on the other hand, exhibit bilateral symmetry around the QC metric.
LIME results show dense regions skewed towards lower QC and PA values, especially for models like DistilBERT and ELECTRA. GPT and BERT models exhibit consistent behavior across interpretation methods, with vertical solid clusters in the QC metric, suggesting consistent MC values irrespective of query counts. DistilBERT and ELECTRA models exhibit different patterns. While their SHAP plots show vertical clusters, LIME displays more horizontally aligned behaviors, especially about the PA metric. CANINE, FNet, and XLM-R models have unique interpretations. CANINE displays a symmetrical pattern in the SM plot, while FNet and XLM-R exhibit scattered regions, especially with the LIME interpretation method. In \autoref{fig:mc_qc_pa_news}, The distributions of most models with SHAP on the AG News dataset are vertically stretched, indicating consistent MC across various QC values, particularly in the GPT model.
Patterns in the SM method are more diverse, with BERT and DistilBERT showing a distinct split, particularly with the QC metric, while GPT retains its vertical orientation. Using the LIME methodology, most models, particularly BERT and DistilBERT, tend to shift towards lower values on the QC metric, indicating fewer queries are required for the misclassification. The GPT model displays consistent vertical concentration across interpretation methods, suggesting stable MC regardless of QC or PA. BERT and DistilBERT models exhibit distinct patterns in each interpretation. SHAP and SM have MC distributions aligned vertically, whereas LIME is more horizontal, indicating that PA is the dominant factor. Different interpretation methods produce diverse patterns in ELECTRA, CANINE, FNet, and XLM-R models. For instance, when using LIME on CANINE's data, the PA metric shows a horizontally skewed pattern, while ELECTRA has a slight concentration in lower QC values under the same methodology.

\ed{In \autoref{fig:mc_qc_pa_yahoo}, SHAP-based MC–QC plots for GPT, BERT, and DistilBERT display dense, horizontally stretched clusters, indicating consistent misclassification confidence across a range of query counts. SM plots for these models show compact, circular distributions with moderate spread, while CANINE, FNet, and XLM-R have slightly broader and less defined clusters. Under LIME, GPT and BERT maintain horizontally extended patterns with some vertical spread, suggesting a slight sensitivity to the amount of perturbation.} 
\ed{CANINE, FNet, and XLM-R under LIME show highly scattered patterns with no clear alignment, indicating significant instability in model responses during attack.}

After conducting a thorough analysis, it has become evident that \ours on different machine learning models, as evaluated through the AG News dataset, has revealed a diverse scope of model behaviors. It is worth noting that specific models exhibit consistent behaviors where their attack outcomes are not significantly affected by input text length (TL), query count (QC), or perturbation amount (PA). Different models show more complex interactions, where the attack's success depends on a combination of multiple metrics. This broader perspective highlights two key points: first, the dataset and method used to interpret data can significantly impact the attack outcomes, and second, understanding these subtle interactions is crucial for fine-tuning attack strategies. Such understanding of vulnerabilities leads to better attack and defense in machine learning.

\section{Discussion} \label{sec:discussion}

\ours has the advantage of its stealthy and potent nature. Its ability to introduce minor changes to the textual input is evident, causing significant shifts in model predictions. Another noteworthy feature is its capability to alter text without compromising its original interpretation, ensuring that adversarial inputs are still plausible to human evaluators. Targeting the most critical tokens strengthens the attack's precision, subtly revealing the model's vulnerabilities.

There are also some challenges for the attack to be used. Advanced input detection techniques can potentially flag the simple character-level modifications employed by the attack. Additionally, with more layers of defense in the INLPS, executing the attack becomes much more complicated. The attack's dependence on determining the most critical tokens might make it less effective against models that distribute importance more evenly across tokens.

Given the challenges posed by the attack, we recommend a few strategies to counter its effects: (1) implementing input sanitization as a frontline defense,  (2) improving the model's interpretability to assist in tracing its reasoning processes, and (3) adversarial training to help the models' robustness. 

\begin{table}[]
\caption{The attack success rate of three models trained on the SST-2 dataset, including adversarial symbols.}
\label{tab:defense}
\centering
\begin{tabular}{c|c|c|c}
\toprule
\textbf{Models} & \textbf{GPT-2} & \textbf{BERT}  & \textbf{DistilBERT} \\ \midrule
\textbf{ASR}    & 0.25 & 0.24 & 0.25      \\ \bottomrule
\end{tabular}
\end{table}

To check the effectiveness of adversarial training defense, we assessed how well three different models - GPT-2, BERT, and DistilBERT could stand up against our attack. We trained the models for a short time on the SST-2 dataset, to which we injected some adversarial symbols into inputs randomly. 
We used the ASR metric to check how often the attack is successful. BERT performed a bit better than the others, with a lower ASR of 0.24, while GPT-2 and DistilBERT both had an ASR of 0.25 (see \autoref{tab:defense}). Compared with the results in \autoref{tab:classifier_result}, we can see an improvement in the robustness of the models against this type of attack.

\section{Conclusion} \label{sec:conc}

Adversarial attacks such as \ours demonstrate that there are creative methods to deceive machine learning systems. During our research, we discovered that \ours can manipulate the text by making minor changes, leading to significant errors in how an NLP system interprets it. This is especially true when the system is not aware of the precise nature of the attack. We also pointed out some cases where \ours might not work. Knowing these weak spots helps us think about better attacks in the future and, more importantly, ways to defend against them. Overall, our work on \ours reminds us that while there are always new challenges in machine learning, there are also new solutions. We hope AI systems will become more innovative and safer as we continue exploring.

\balance
\bibliography{ref.bib}
\bibliographystyle{IEEEtran}

\end{document}